\documentclass[runningheads]{llncs}
\usepackage{graphicx}
\usepackage{hyperref}

\usepackage{amsmath}
\usepackage{listings}
\usepackage{color}

\usepackage{microtype}
\usepackage{wrapfig}

\usepackage{booktabs}
\usepackage{bbding}

\usepackage[font=footnotesize]{caption}

\usepackage[]{algorithm2e}

\usepackage{xcolor}
\hypersetup{
    colorlinks,
	linkcolor={blue}, 
	citecolor={blue}, 
	urlcolor={blue}   
}

\usepackage[misc]{ifsym}

\definecolor{codegreen}{rgb}{0,0.6,0}
\definecolor{codegray}{rgb}{0.5,0.5,0.5}
\definecolor{codepurple}{rgb}{0.58,0,0.82}
\definecolor{backcolour}{rgb}{0.95,0.95,0.92}
 
\lstdefinestyle{mystyle}{
    commentstyle=\color{codegreen},
    keywordstyle=\color{magenta},
    numberstyle=\tiny\color{codegray},
    stringstyle=\color{codepurple},
    basicstyle=\footnotesize,
    breakatwhitespace=false,         
    breaklines=true,                 
    captionpos=b,                    
    keepspaces=true,                 
    numbersep=5pt,                  
    showspaces=false,                
    showstringspaces=false,
    showtabs=false,                  
    tabsize=2
}

\newsavebox{\imagebox}

\usepackage{cleveref}
\usepackage{subfig}
\usepackage[export]{adjustbox}

\begin{document}
\title{Active-Code Replacement in the \\OODIDA Data Analytics Platform\thanks{The final authenticated version is available online at \url{https://doi.org/10.1007/978-3-030-48340-1_55}.}} 
\titlerunning{Active-Code Replacement in the OODIDA Data Analytics Platform}
%


\author{Gregor Ulm\inst{1, 2}\Envelope
\orcidID{0000-0001-7848-4883}
\and
Emil Gustavsson \inst{1, 2}
\orcidID{0000-0002-1290-9989}
\and
Mats Jirstrand \inst{1, 2}
\orcidID{0000-0002-6612-8037}
}
\authorrunning{G. Ulm et al.}

\institute{Fraunhofer-Chalmers Research Centre for Industrial Mathematics,\\ Chalmers Science
Park, 412 88 Gothenburg, Sweden\\
\and Fraunhofer Center for Machine Learning,
\\Chalmers Science
Park, 412 88 Gothenburg, Sweden\\
\email{\{gregor.ulm, emil.gustavsson, mats.jirstrand\}@fcc.chalmers.se}\\
\url{http://www.fcc.chalmers.se/}
}
\maketitle              
\begin{abstract}

OODIDA (On-board/Off-board Distributed Data Analytics) is a platform for distributing and executing concurrent data analytics tasks. It targets fleets of reference vehicles in the automotive industry and has a particular focus on rapid prototyping. Its underlying message-passing infrastructure has been implemented in Erlang/OTP. External Python applications perform data analytics tasks. Most work is performed by clients (on-board). A central cloud server performs supplementary tasks (off-board). OODIDA can be automatically packaged and deployed, which necessitates restarting parts of the system, or all of it. This is potentially disruptive. To address this issue, we added the ability to execute user-defined Python modules on clients as well as the server. These modules can be replaced without restarting any part of the system and they can even be replaced between iterations of an ongoing assignment. This facilitates use cases such as iterative A/B testing of machine learning algorithms or modifying experimental algorithms on-the-fly.

\end{abstract}
\section{Introduction}
OODIDA is a modular system for concurrent distributed data analytics for the automotive domain, targeting fleets of reference vehicles~\cite{ulm2019oodida}. Its main purpose is to process telemetry data at its source as opposed to transferring all data over the network and processing it on a central cloud server (cf.~Fig.~\ref{fig:whole_fleet}). A data analyst interacting with this system uses a Python library that assists in creating and validating assignment specifications. Updating this system with new computational methods necessitates terminating and redeploying software. However, we would like to perform updates without terminating ongoing tasks. We have therefore extended our system with the ability to execute user-defined code both on client devices (on-board) and the cloud server (off-board), without having to redeploy any part of it. As a consequence, OODIDA is now highly suited for rapid prototyping. The key aspect of our work is that active-code replacement of Python modules piggybacks on the existing Erlang/OTP infrastructure of OODIDA for sending assignments to clients, leading to a clean design. This paper is a condensed version of a work-in-progress paper~\cite{ulm2019active}, giving an overview of our problem (Sect.~\ref{problem}) and its solution (Sect.~\ref{solution}), followed by an evaluation (Sect.~\ref{eval}) and related work~(Sect.~\ref{related}). 

\vspace{-2.3em}
\newsavebox{\tempbox}
\begin{figure}
\sbox{\tempbox}{\includegraphics[width=0.45\textwidth, trim=2cm 2.5cm 2cm 2.5cm, clip]{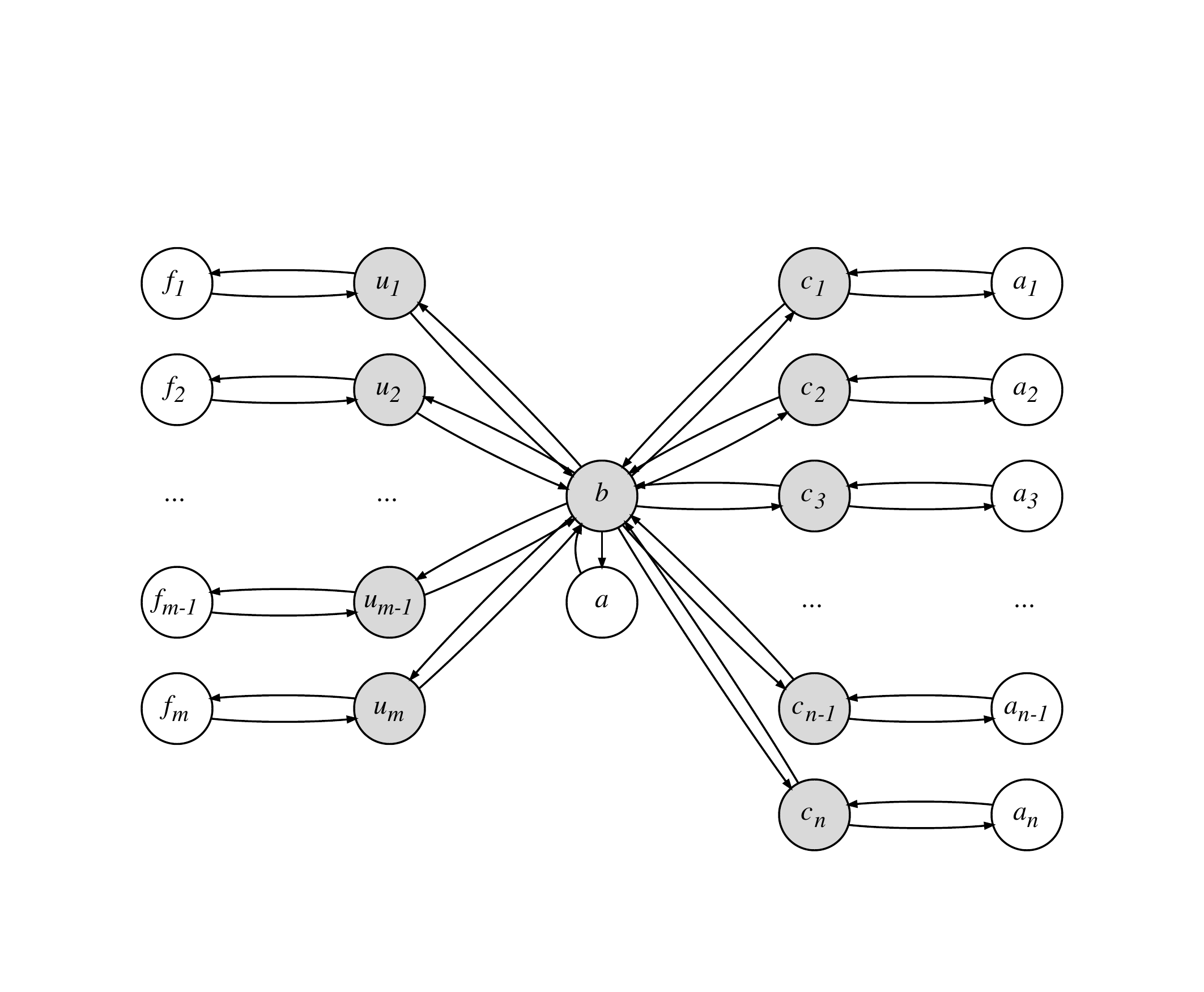}}
\subfloat[OODIDA in context]{\usebox{\tempbox}\label{fig:context}}%
\qquad
\subfloat[Whole-fleet assignment]{\vbox to \ht\tempbox{%
  \vfil
  \hbox to 0.45\textwidth{
  \includegraphics[width=0.45\textwidth, trim=2cm 2.5cm 2cm 2.5cm, clip]{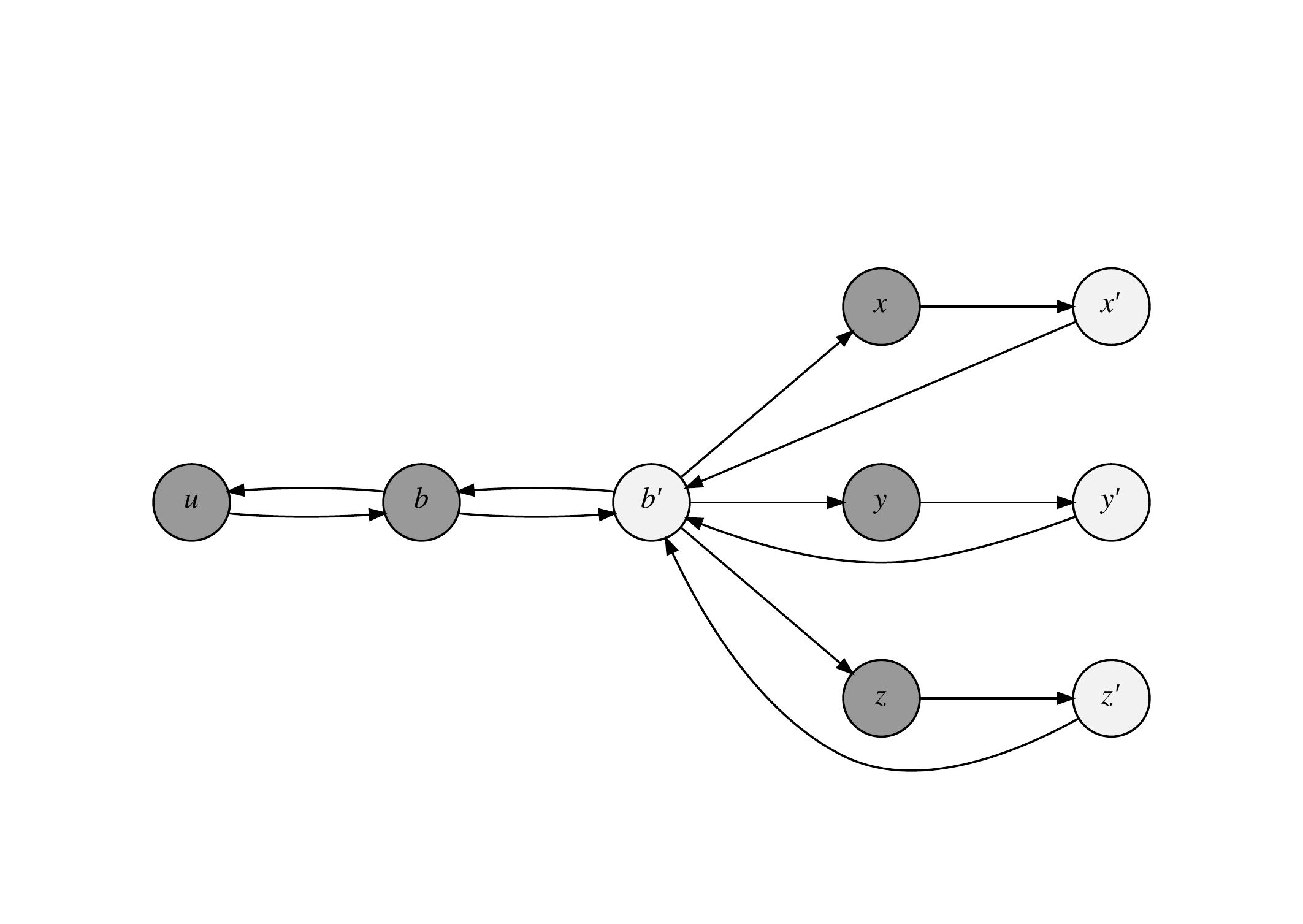}
  }
  \vfil}\label{fig:whole_fleet_assignment}}%

\caption{OODIDA overview and details: In (a) user nodes $\boldsymbol{u}$ connect to a central cloud $b$, which connects to clients $\boldsymbol{c}$. The shaded nodes are implemented in Erlang/OTP; the other nodes are external Python applications, i.e.\ the user front-ends $\boldsymbol{f}$, the external server application $a$, and external client applications $\boldsymbol{a}$. In (b) the core of OODIDA is shown with permanent nodes (dark) and temporary handlers (light) in an instance of a whole-fleet assignment. Cloud node $b$ spawned an assignment handler $b'$. After receiving an incoming task, clients $x, y$ and $z$ spawned task handlers $x'$, $y'$, and $z'$ that interact with external applications. Nodes $x$ and $x'$ correspond to $c_1$ in (a) etc.}
\label{fig:whole_fleet}
\end{figure}

\section{Problem}
\label{problem}
OODIDA has been designed for rapid prototyping, which implies that it frequently needs to be extended with new computational methods, both for on-board and off-board data processing. To achieve this goal, Python applications on the cloud and clients have to be updated. Assuming that we update both, the following steps are required: The user front-end $f$ needs to be modified to recognize the new off-board and on-board keywords for the added methods, including checks of assignment parameter values. In addition, the cloud and client applications have to be extended with the new methods. All ongoing assignments need to be terminated and the cloud and clients shut down. Afterwards, we can redeploy and restart the system. This is disruptive, even without taking into account potentially long-winded software development processes in large organizations. On the other hand, the turn-around time for adding custom methods would be much shorter if we could do so at runtime. Active-code replacement targets this particular problem, with the goal of further improving the suitability of OODIDA for rapid prototyping.

\section{Solution}
\label{solution}
With active-code replacement, the user can define a custom Python module for the cloud and for client devices. It is implemented as a special case of an assignment. The front-end $f$ performs static and dynamic checks, attempting to verify correctness of syntax and data types. If these checks succeed, the provided code is turned into a JSON object and ingested by user node $u$ for further processing. Within this JSON object, the user-defined code is stored as an encoded text string. It is forwarded to cloud node $b$, which spawns an assignment handler $b'$ for this particular assignment. Custom code can be used on the cloud and/or clients. Assuming clients have been targeted with active-code replacement, node $b'$ turns the assignment specification into tasks for all clients $\boldsymbol{c}$ specified in the assignment. Afterwards, task specifications are sent to the specified client devices. There, the client process spawns a task handler for the current task, which monitors task completion. The task handler sends the task specification in JSON to an external Python application, which turns the given code into a file, thus recreating the Python module the data analyst initially provided. The resulting files are tied to the ID of the user who provided it. After the task handler is done, it notifies the assignment handler $b'$  and terminates. Similarly, once the assignment handler has received responses from all task handlers, it sends a status message to the cloud node and terminates. The cloud node sends a status message to inform the user that their custom code has been successfully deployed. Deploying custom code to the cloud is similar, the main difference being that $b'$ communicates with the external Python application on the cloud.

If a custom on-board or off-board computation is triggered by a special keyword in an assignment specification, Python loads the user-provided module. The user-specified module is located at a predefined path, which is known to the Python application. The custom function is applied to the available data after the user-specified number of values has been collected. When an assignment uses custom code, external applications reload the custom module with each iteration of an assignment. This leads to greater flexibility: Consider an assignment that runs for an indefinite number of iterations. As external applications can process tasks concurrently, and code replacement is just another task, the data analyst can react to intermediate results of an ongoing assignment by deploying custom code with modified algorithmic parameters while this assignment is ongoing. As custom code is tied to a user ID, there is furthermore no interference due to custom code that was deployed by other users. The description of active-code replacement so far indicates that the user can execute custom code on the cloud server and clients, as long as the correct inputs and outputs are consumed and produced. What may not be immediately obvious, however, is that we can now create \emph{ad hoc} implementations of even the most complex OODIDA use cases in custom code, such as federated learning~\cite{mcmahan2016communication}.

Inconsistent updates are a problem in practice, i.e.~results sent from clients may have been produced with different custom code modules in the same iteration of an assignment. This happens if not all clients receive the updated custom code before the end of the current iteration. To solve this problem, each provided module with custom code is tagged with its md5 hash signature, which is reported together with the results from the clients. The cloud only uses the results tagged with the signature that achieves a majority. Consequently, results are never tainted by using different versions of custom code in the same iteration.

\section{Evaluation}
\label{eval}

The main benefit of active-code replacement is that code for new computational methods can be deployed right away and executed almost instantly, without affecting other ongoing tasks. In contrast, a standard update of the cloud or client installation necessitates redeploying and restarting the respective components of the system. In an idealized test setup, where the various workstations that run the user, cloud and client components of OODIDA are connected via Ethernet, it takes a fraction of a second for a custom on-board or off-board method to be available for the user to call when deployed with active-code replacement, as shown in Table~\ref{tab:comparison}. On the other hand, automated redeployment of the cloud and client installation takes roughly 20 and 40 seconds, respectively. The runtime difference between a standard update and active-code replacement amounts to three orders of magnitude. Of course, real-world deployment via a wireless or 4G connection would be slower as well as error-prone. Yet, the idealized evaluation environment reveals the relative performance difference of both approaches, eliminating potentially unreliable data transmission as a source of error.

This comparison neglects that, compared to a standard update, active-code replacement is less bureaucratic and less intrusive as it does not require interrupting any currently ongoing assignments. Also, in a realistic industry scenario, an update could take days or even weeks due to software development and organizational processes. However, it is not the case that active-code replacement fully sidesteps the need to update the library of computational methods on the cloud or on clients as OODIDA enforces restrictions on custom code. For instance, some parts of the Python standard library are off-limits. Also, the user cannot install external libraries. Yet, for typical algorithmic explorations, which users of our system regularly conduct, active-code replacement is a vital feature that increases user productivity far more than the previous comparison may imply. That being said, due to the limitations of active-code replacement, it is complementary to the standard update procedure rather than a competitive approach.

\begin{table}[]
\centering{
\caption{Runtime comparison of active-code replacement of a moderately long Python module versus regular redeployment in an idealized setting. The former has a significant advantage. Yet, this does not factor in that a standard update is more invasive but can also be more comprehensive. The provided figures are the averages of five runs.\\}
\label{tab:comparison}
\begin{tabular}{@{}lll@{}}
\toprule
& Cloud & Client \\ \toprule
Active-code replacement \phantom{aaaaaaa}
&  20.3 ms   \phantom{aaaaaaa} & 45.4 ms        \\
Standard redeployment
& 23.6 s  &  40.8 s\\
\bottomrule
\end{tabular}
}
\end{table}

\section{Related Work}
\label{related}
The feature described in this paper is an extension of the OODIDA platform~\cite{ulm2019oodida}, which originated from \texttt{ffl-erl}, a framework for federated learning in Erlang/OTP~\cite{ulm2019b}. In terms of descriptions of systems that perform active-code replacement, Polus by Chen et al.~\cite{chen2007polus} deserves mention. A significant difference is that it replaces larger units of code instead of isolated modules. It also operates in a multi-threading environment instead of the highly concurrent message-passing environment of OODIDA. We also noticed a similarity between our approach and Javelus by Gu et al.~\cite{gu2012javelus}. Even though they focus on updating a stand-alone Java application as opposed to a distributed system, their described "lazy update mechanism" likewise only has an effect if a module is indeed used. This mirrors our approach of only loading a custom module when it is needed.


\subsubsection*{Acknowledgements.} 
\small{
This research was financially supported by the project On-board/Off-board Distributed Data Analytics (OODIDA) in the funding program FFI: Strategic Vehicle Research and Innovation (DNR 2016-04260), which is administered by VINNOVA, the Swedish Government Agency for Innovation Systems. It took place in the Fraunhofer Cluster of Excellence "Cognitive Internet Technologies." Simon Smith and Adrian Nilsson helped with a supplementary part of the implementation of this feature and carried out the performance evaluation. Ramin Yahyapour (University of G{\"o}ttingen) provided insightful comments during a poster presentation.
}
%
%
%
\bibliographystyle{splncs04}

\end{document}